# LENSING BY DISTANT CLUSTERS: HST OBSERVATIONS OF WEAK SHEAR IN THE FIELD OF 3C324[*]

Ian Smail[1] and Mark Dickinson[2]

1) The Observatories of the Carnegie Institution of Washington, 813 Santa Barbara St., Pasadena, CA 91101-1292

2) Space Telescope Science Institute, 3700 San Martin Dr, Baltimore, MD 21218.

## ABSTRACT

We present the detection of weak gravitational lensing in the field of the radio galaxy 3C324 ($z = 1.206$) using deep HST imaging. From an analysis of the shapes of faint $R = 24.5$–$27.5$ galaxies in the field we measure a weak, coherent distortion centered close to the radio source. This shear field most likely arises from gravitational lensing of distant field galaxies by a foreground mass concentration. In the light of previous observations of this region, which indicate the presence of a rich cluster around the radio source, we suggest that the most likely candidate for the lens is the cluster associated with the radio galaxy at $z = 1.2$. If so, this is the most distant cluster to have been detected by weak shear observations. Such a statement has two important consequences. Firstly, it shows that massive, collapsed structures exist in the high redshift Universe, and secondly that a significant fraction of the $R = 24.5$–$27.5$ field galaxy population lies beyond $z \sim 1.2$.

*Subject headings:* cosmology: observations – galaxies: evolution – clusters: general – gravitational lensing.

---

## 1. Introduction

Observations of weak gravitational lensing are a relatively new tool to study the distribution of mass in the distant Universe. A growing number of papers employ this technique to investigate mass distributions on a range of scales, from individual galaxies (Brainerd, Blandford & Smail 1995), through groups and poor clusters (Fort et al. 1995), rich clusters (Tyson, Valdes & Wenk 1990, Smail et al. 1995a, Squires et al. 1995, Kneib et al. 1995) and onto larger scale structures (Mould et al. 1994). Lensing is particularly useful in this regard because it does not require the mass in the lens to be virialised on the scales of interest. Lensing can also be used to constrain the redshifts of the very faint objects viewed through rich clusters by inverting simple geometrical models of the lens (Kneib et al. 1995). The primary limitation of this technique when applied to $z \lesssim 0.5$ clusters is the difficulty in obtaining accurate redshifts for the most distant objects, $z \gtrsim 1.5$. This problem can be circumvented if we can find a lens which is itself distant, $z \sim 1$ (c.f. Smail, Ellis & Fitchett 1994).

This letter presents one application of the weak lensing method to investigate the shear field in the vicinity of the distant radio galaxy 3C324 ($z = 1.206$) imaged with the *Hubble Space Telescope* (HST). HST is a unique instrument for this type of study providing accurate shapes for faint galaxies, free of the systematic effects which hamper ground-based applications. We can thus expect to be able to measure weak shear signals at levels ($\lesssim 1$–3%) which are difficult to achieve on ground-based telescopes.

We report a statistically significant shear pattern around 3C324. Previous observations (reviewed below) have indicated that 3C324 lies in a rich cluster of galaxies (Spinrad & Djorgovski 1984, Dickinson 1994, 1995a). We suggest on the basis of these data that the likely source of the observed shear is the cluster at $z = 1.2$ associated with the radio source. We then present joint constraints on the properties of the lensing cluster and the distant field population which can be drawn from these observations.

## 2. Observations, Analysis and Results

### 2.1. Observations and Reduction

The field around 3C324 was observed with WFPC2 on-board HST for a total of 32 orbits in 5 visits between 11 May and 15 June 1994. These deep images were obtained in order to study the morphologies of the faint, cluster galaxies (Dickinson et al. 1995). The final on-source integration time was 64.8 ksec, all in the F702W ($R$) filter. The individual images were acquired at the same roll angle with the same guide stars, but with small offsets between sets of observations. Each WFPC2 exposure produces four separate frames, three for the Wide Field Camera (WFC) and one for the Planetary Camera (PC). These were mosaiced together using a transformation which accounts for the geometric distortion of the WFPC2 optics. The mosaics were then aligned and combined with a clipping algorithm to reject cosmic ray events and other pixel defects. All reductions were carried out using standard IRAF/STSDAS routines. The stacked exposure is shown in Figure 1 (Plate XX).

To catalog the faint galaxies within this frame we use the SExtractor image analysis package (Bertin 1994). This software uses an isophotal area criterion to select objects detected above a fixed surface brightness threshold. It then provides photometry and shape information for all deblended objects. For this analysis we adopted a minimum area, after smoothing with a 0.3 arcsec diameter top-hat filter, of at least 0.12 arcsec$^2$ above the $\mu_R = 26.4$ mag arcsec$^{-2}$ isophote, equivalent to 1.0$\sigma$ per pixel. To convert from instrumental magnitudes to $R$ we take the conversion of Holtzman et al. (1995) and use $(V–R) \sim 0.5$ (Smail et al. 1995b).

The 3C324 field has a 5$\sigma$ detection limit of $R = 27.7$ within our 1.0 arcsec diameter photometry aperture, and contains $\sim 1040$ objects to this limit in the 4.8 arcmin$^2$ of the 3 WFC chips. We discard both the lower sensitivity PC chip and also the borders of the WFC chips from our analysis. We apply three cuts to this catalog before using it in the lensing analysis. The first two cuts are on magnitude. We discard all objects fainter than $R = 27.5$ to remove objects with inadequate signal-to-noise for measuring reliable shapes. We also apply a bright magnitude cut to reduce foreground contamination, retaining only galaxies with $R > 24.5$, approximately 1 magnitude fainter than the brightest galaxies in $z \sim 1$ clusters. Finally, in order to remove objects which are too round to accurately determine their orientations, we remove all objects with ellipticities $\epsilon < 0.05$, where $\epsilon = (a-b)/(a+b)$ where $a$ and $b$ are the major and minor axis lengths of an object. The final catalog used in the lensing analysis contains $\sim 700$ objects with a surface density of $\sim 145$ arcmin$^{-2}$.

## 2.2. Analysis and Results

To search for the effects of lensing we evaluate the complex shear: $g = \epsilon e^{2i\phi}$, where $\phi$ is the position angle of the major axis of the object, for every object in the catalog (Kneib et al. 1995). These shear estimates are then binned into independent $20'' \times 20''$ cells and plotted as a vector field over the image in Figure 1. Although a visual inspection of the image shows no good candidates for strongly lensed features, it is possible with some imagination to pick out a weak circular pattern in the shear vectors around the radio galaxy. To show this better we use the method of Kochanek (1990) to determine a center for the whole shear field. Each vector defines a probability distribution for the likely center, in the form of two cones extending perpendicular to the shear direction with opening angles determined from a combination of the error in the shear direction, estimated from boot-strap resampling, and the shear strength at that point. The maximum of the combined probability distributions from all the vectors then gives the best center for the whole shear field. The likelihood distribution for the center is shown by probability contours plotted on Figure 1. As suspected, the shear field is centered close to the radio galaxy, with its peak lying $\sim 100 h^{-1}$ kpc[†] to the south-west. In the remainder of the analysis we have adopted the radio source as the center of the lens.

Having determined that a coherent shear field exists around the radio source, we next wish to measure its strength and radial profile. We first calculate the average tangential shear ($g_1 = -\epsilon \cos(2\theta)$, where $\theta$ is the angle between an object's major axis and the vector joining it to the lens center) within an annulus from $r = 50-250 h^{-1}$ kpc (10-60 arcsec) around the radio source and obtain $<g_1> = 0.031 \pm 0.010$. Figure 2 plots $g_1$ averaged in circular annuli around 3C324. The mean tangential shear drops by roughly a factor of 5 between $50-400 h^{-1}$ kpc, similar to the decline expected from lensing by a spherical, isothermal mass distribution.

To assess possible systematic effects from the telescope optics, detectors, or the reduction method, we have applied the same analysis to a similarly deep HST image. This archival data (J. Westphal, P.I.) consists of deep $V$ and $I$-band WFPC-2 images of the deepest field within the "Groth Strip," a blank field survey which does not target any known cluster. The data was processed and object catalogs were generated in a similar manner as for the 3C324 dataset. Although the total exposure time is somewhat shorter, the galaxy surface density in this field is comparable to that available in the 3C324 observation. Within an annulus of 10-60 arcsec around the equivalent position of 3C324 on the PC, we detect no significant tangential distortion ($<g_1> = -0.002 \pm 0.012$). We tentatively conclude that the observed shear in the 3C324 field is neither instrumental in origin, nor a result of our reduction methods.

---

[†] We take $q_\circ = 0.5$ and $h = H_\circ/100$ km/sec/Mpc, which gives $4.30 h^{-1}$ kpc/arcsec at $z = 1.206$

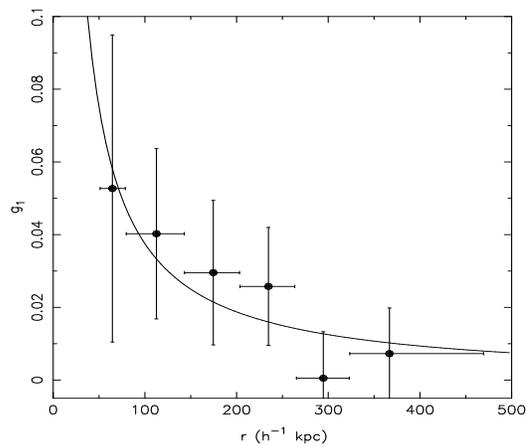

**Figure 2.** The radial dependency of the tangential shear strength ($g_1$) around 3C324. The error bars are $1\sigma$ limits from boot-strap resampling of the data. We also plot a $g_1 \propto r^\alpha$ line with $\alpha = -1$, showing the expected decline in signal if it arises from lensing by a spherical, isothermal mass distribution. This decline is formally consistent with the observations, which give $\alpha = -1.0 \pm 0.2$.

## 3. Discussion

We have demonstrated that a coherent shear pattern exists in the field of 3C324 and that this pattern is apparently not an artifact of the telescope optics. Moreover, the lensing pattern is centered close to the radio galaxy and shows a radial profile which is consistent with that seen from lensing by moderate and intermediate redshift clusters (Fort & Mellier 1994). The available evidence therefore points towards the presence of a substantial mass concentration in the field of 3C324, and moreover this mass concentration lies in front of a sizeable fraction of the $R = 24.5-27.5$ field population. We now address the question of the likely redshift of the lens.

There is a marked excess of faint, red galaxies con-centrated around 3C324 (Dickinson 1995a, 1995b). In the $K$-band the overdensity relative to faint field number counts reaches a factor of $\sim 6$. Many of these galaxies have E/S0 morphologies in the WFPC2 image and lie in a narrow color range with $(R-K) \approx 6$, too red for ordinary, unreddened galaxies at $z < 1$. Spectroscopic observations of these galaxies is currently underway (Dickinson et al. 1995). In the remainder of our discussion we will assume that the shear field derives solely from a distant cluster, probably associated with the radio source at $z \approx 1.2$. Here we rely upon the proximity of the radio galaxy to the center of the shear field to identify the 3C324 cluster as the lens. This is an important assumption: although the measured shear is statistically significant, it is also sufficiently weak that it could arise from a modest mass concentration at a lower redshift (c.f. Le Fèvre et al. 1987, Fort et al. 1995).

The observed shear strength depends jointly upon the distances of the background lensed galaxies and the mass within the lens. We will therefore investigate the allowed ranges for these two parameters by assuming plausible values for one parameter and determining the most likely value of the other. We start by examining the mass predictions.

In view of the low signal strength and restricted field of view, complex modeling is not warranted. For simplicity, we will describe the lens by a spherical, singular isothermal mass profile with a single parameter: the velocity dispersion, $\sigma_{cl}$. We also take the following functional form (Efstathiou et al. 1990) to parametrise the redshift distribution, $N(z)$, of the $R = 24.5$–$27.5$ field population:

$$N(z) \propto z^2 e^{-(z/z_0)^\beta}.$$

For our fiducial model, we adopt $\beta = 5$, for which $<z> = 0.78 z_0$. We focus on this model because it is a reasonable description of the $R \sim 25$ field $N(z)$ recently derived by Kneib et al. (1995). Those authors have inverted a well-constrained mass model of the rich cluster A2218 to estimate redshifts for $\sim 60$ distorted arcs and arclets in the core. Their sample reached $R \sim 25$, at which point they obtain $<z> \sim 0.8$, with $\sim 20\%$ of the population at $z \gtrsim 1$. Extrapolating this result to the magnitude limit of the present study predicts $<z> \sim 1.2$. Using our adopted form for $N(z)$, Figure 3a plots the average tangential shear expected within our aperture as a function of $\sigma_{cl}$. The three curves show the dependence on the assumed mean redshift for the $R = 24.5$–$27.5$ field population, using $<z> = 1.0$, 1.2 and 1.5. These place 25%, 50% and 75% (respectively) of the faint galaxies at $z > 1.2$, and are so labeled. The high redshift of the lens, comparable to the mean redshift of the background sources, means that the mass estimate is very sensitive to the exact choice for $N(z)$. For the fiducial "no evolution" model with $<z> \sim 1.2$, the observed shear corresponds to a velocity dispersion $\sigma_{cl} \sim 1300^{+180}_{-230}$ km sec$^{-1}$, equivalent to a mass of $M(r < 250 h^{-1} \text{kpc}) \sim 3 \times 10^{14} h^{-1} M_\odot$.

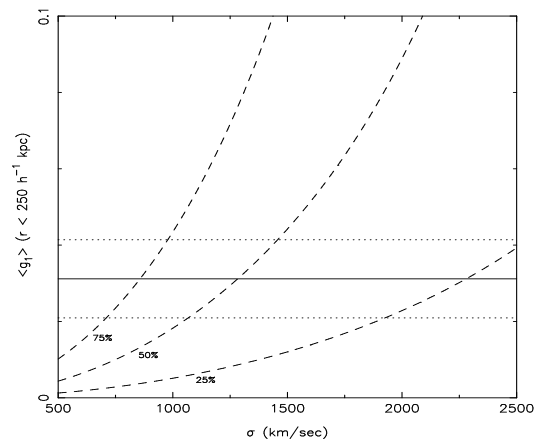

**Figure 3 (a).** The relationship between the average tangential shear, $<g_1>$, measured in an annulus from 50–$250 h^{-1}$ kpc, and the lens mass, parameterised by $\sigma_{cl}$. The functional form used for the galaxy redshift distribution is given in the text and we adopt $\beta = 5$ and an 80% efficiency for the shear measurement (Smail et al. in prep.). The three dashed curves show the predictions using different mean redshifts for the field population, labeled by the fraction of the population lying beyond $z = 1.2$. The solid line shows the measured shear, with $\pm 1\sigma$ bounds on this value shown by dotted lines. We derive $\sigma_{cl} = 1300^{+180}_{-230}$ km sec$^{-1}$ using the "no-evolution" model (50% with $z > 1.2$).

It is clear from Figure 3a that for any plausible lens mass a substantial fraction ($\gtrsim 25\%$) of the faint field population must lie beyond $z \sim 1.2$. To further quantify this statement, we use an independent measure of the cluster mass and ask what is the shallowest faint galaxy $N(z)$ which is consistent with the observed shear signal. Here, we make use of ROSAT observations of this field (Dickinson & Mushotzky 1995). An X-ray source coincident with 3C324 was initially detected in a 15.4 ksec ROSAT PSPC exposure, and was shown to be spatially extended using a deep 72.1 ksec integration with the HRI. Assuming that the entire flux arises from a hot intra-cluster plasma at the radio galaxy redshift with a temperature of 5 keV, metallicity 0.5 × solar, and cosmic element abundance ratios, Dickinson & Mushotzky (1995) determine a bolometric X-ray luminosity of $L_X = (2.0 \pm 0.4)h^{-2} \times 10^{44}$ erg sec$^{-1}$.

For a nearby cluster with this luminosity, the local $L_X$-$\sigma_{cl}$ relationship (Edge & Stewart 1990) would imply a rest-frame velocity dispersion of $\sigma_{cl} \sim 900^{+430}_{-300}$ km sec$^{-1}$. In view of possible differential evolution between the X-ray and optical properties of rich clusters observed to $z \sim 0.5$ (Castander et al. 1995), this may actually be a lower limit to the most likely value. Turning to Figure 3b, we find that a cluster with this $\sigma_{cl}$ would require a mean redshift $<z> \sim 1.5^{+0.8}_{-0.3}$ for the $R = 24.5$–27.5 field population. Equivalently, at least $70 \pm 20\%$ of the population lies at $z \gtrsim 1.2$, marginal greater than the 50% expected from the no-evolution model. Either way, it appears that at least half of the $R \sim 27$ field population is at redshifts $z \gtrsim 1$.

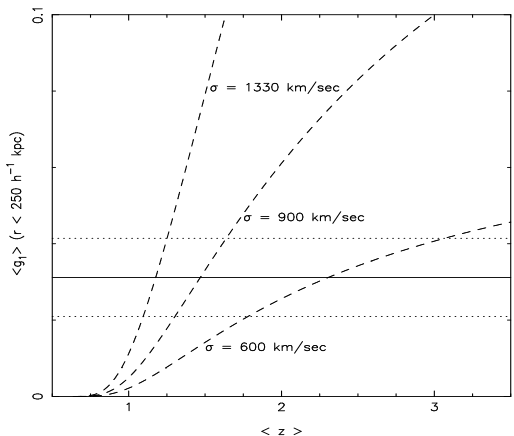

**Figure 3 (b).** The variation in $<g_1>$ with the mean redshift, $<z>$, of the faint field population. The three dashed curves trace the bounds on $\sigma_{cl}$ as estimated from the observed X-ray luminosity of the cluster and the local $L_X$-$\sigma_{cl}$ relationship: $\sigma_{cl} \sim 900^{+430}_{-300}$ km sec$^{-1}$. These translate into $<z> = 1.5^{+0.8}_{-0.3}$ for the galaxy population with $24.5 < R < 27.5$.

This conclusion has important consequences for models of faint galaxy evolution. In particular, some recent models have proposed that the very faint field population is dominated by dwarf galaxies at $z < 1$. For example, Babul & Ferguson (1995) consider a scenario where the faint counts are dominated by a multitude of "bursting dwarfs," whose formation is delayed until $z < 1$. This model adequately reproduces observed galaxy counts and sizes, as well as their redshift distributions to the limits of existing spectroscopic surveys. At the fainter magnitudes probed by this lensing experiment, however, the model predicts much lower mean redshifts, with only $\sim 10\%$ of the $24.5 < R < 27.5$ galaxies lying at $z > 1.2$ (H. Ferguson, private communication).

## 4. Conclusions

• We present a detection of weak shear in the field of the distant radio galaxy 3C324 at $z = 1.206$. We conclude that the shear arises from a mass concentration in the direction of the radio galaxy, and appeal to other observations of this field to suggest that the shear is produced by a cluster associated with the radio source at $z = 1.2$.

• Simple modelling of the lensing cluster as an isothermal mass distribution provides joint constraints on the redshift distribution of the faint field population to $R = 27.5$ and the amount of mass in the lensing cluster. Adopting the no-evolution $N(z)$ for our magnitude limits we predict $<z> \sim 1.2$. Taking this $N(z)$ we estimate a velocity dispersion for the cluster of $\sigma_{cl} \sim 1300^{+180}_{-230}$ km sec$^{-1}$ from the observed shear signal, although this value is very sensitive to the exact choice of $N(z)$. At a redshift of $z \sim 1.2$ collapsed structures of this mass are predicted to be extremely rare in standard cosmological models (Jing & Fang 1994).

• Alternatively, we can fix the mass of the lens and determine the mean redshift of the faint galaxies in our sample. If the observed X-ray luminosity of 3C324 derives from a hot intra-cluster medium, and if the mass distribution is reasonably described by an isothermal sphere (an assumption supported by the radial dependence of the shear amplitude), then the local $L_X$-$\sigma_{cl}$ relationship would predict $\sigma_{cl} \sim 900$ km sec$^{-1}$. The observed shear strength would then require a mean redshift for the $R = 24.5$–27.5 field population of $<z> \sim 1.5^{+0.8}_{-0.3}$, with $\sim 70\%$ lying beyond $z \sim 1.2$. We therefore conclude that a large fraction ($\gtrsim 50\%$) of the $R \sim 27$ field population must lie at $z \gtrsim 1$.

## Acknowledgements

We thank Roger Blandford, Richard Ellis, Harry Ferguson, David Hogg and Jean-Paul Kneib for many useful and entertaining discussions. MD thanks his collaborators in the HST imaging program, Adam Stanford, Peter Eisenhardt, Hyron Spinrad, and George Djorgovski. IRS gratefully acknowledges support from a Carnegie Fellowship and through NASA/STScI grant GO-5395. MD similarly acknowledges support from an AURA Fellowship at STScI and through NASA/STScI grant GO-5465.

# REFERENCES

Babul, A., and Ferguson, H.C. 1995, ApJ, submitted.

Bertin E., 1994, SExtractor manual, IAP, Paris.

Brainerd, T.G., Blandford, R.D. & Smail, I., 1995, ApJ, submitted.

Castander, F.J., Bower, R.G., Ellis, R.S., Aragón-Salamanca, A., Mason, K.O., Hasinger, G., McMahon, R.G., Carrera, F.J., Mittaz, J.P.D., Pérez-Fournon, I. & Lehto, H.J., 1995, Nature, in press.

Dickinson, M., 1994, Ph.D. thesis, University of California, Berkeley.

Dickinson, M., 1995a, in "Fresh Views on Elliptical Galaxies", ASP Conference Series, eds. Buzzoni, A., Renzini, A. & Serrano, A.

Dickinson, M., 1995b, in "Galaxies in the Young Universe," ed. H. Hippelein, Springer Verlag.

Dickinson, M., Dey, A., Spinrad, H., Stanford, S.A., Eisenhardt, P.R.M. & Le Fèvre, O., 1995 in preparation.

Dickinson, M. & Mushotzky, R., 1995, in preparation.

Edge, A.C. & Stewart, G.C., 1990, MNRAS, 252, 428.

Efstathiou, G.P., Bernstein, G., Katz, N., Tyson, J.A., Guhathakurta, P., 1990, ApJL, 380, L47.

Le Fevre, O., Hammer, F., Nottale, L. & Mathez, G., 1987, Nature, 326, 268.

Fort, B. & Mellier, Y., 1994, Astron. Astrophys. Rev., 5, 239.

Fort, B., Mellier, Y., Dantel-Fort, M., Bonnet, H. & Kneib, J.-P., 1995, A&A, submitted.

Holtzman, J.A., Burrows, C.J., Casterno, S., Hester, J.J., Trauger, J.T., Watson, A.M. & Worthey, G., 1995, preprint.

Jing, Y.P. & Fang, L.Z., 1994, ApJ, 432, 438.

Kneib, J.-P., Ellis, R.S., Smail, I., Couch, W.J. & Sharples, R.M., 1995, ApJ, submitted.

Kochanek, C.S., 1990, MNRAS, 247, 135.

Mould, J.R., Blandford, R.D., Villumnsen, J.V., Brainerd, T.G., Smail, I., Small, T.A. & Kells, W., 1994, MNRAS, 271, 31.

Smail, I., Ellis, R.S. & Fitchett, M.J., 1994, MNRAS, 270, 245.

Smail, I., Ellis, R.S., Fitchett, M.J. & Edge, A.C., 1995a, MNRAS, 273, 277.

Smail, I., Hogg, D.W., Yan, L. & Cohen, J.L., 1995b, ApJL, 449, L105.

Spinrad, H. & Djorgovski, S.G., ApJ, 285, L49.

Squires, G., Kaiser, N., Babul, A., Fahlman, G., Woods, D., Neumann, D.M. & Böhringer, H., 1995, ApJ, submitted.

Tyson, J.A., Valdes, F. & Wenk, R.A., 1990, ApJL, 349, L1.

# Figures

**Plate 1.** Combined HST WFPC2 image of 3C324 taken through the F702W filter. The radio source lies on the PC chip and is marked. The total exposure time is 64.8 ksec and the effective resolution of this image is 0.1 arcsec. Overlayed on the image are the independent shear vectors determined from the shapes of the faint $R = 24.5\text{--}27.5$ galaxies. The dashed contours indicate the likelihood distribution for the best fitting center of the shear field. The scatter is estimated by repeatedly resampling the shear field and discarding 30% of the vectors. The contours are spaced by factors of 1.5 down from the peak. We show a vector indicating a 10% shear at the upper left.

---

This 2-column preprint was prepared with the AAS LaTeX macros v3.0.